\documentclass[aps,prd,reprint,groupedaddress,showkeys,showpacs]{revtex4-1}
\usepackage{graphicx}
\usepackage{dcolumn}
\usepackage{bm}%
\usepackage{amsmath}
\usepackage[caption = false]{subfig}

\begin{document}

\title{Magnetized taub adiabat and the PT of magnetic neutron stars}

\date{\today}

\author{Ritam Mallick} 
\email{mallick@iiserb.ac.in}
\affiliation{Indian Institute of Science Education and Research Bhopal, Bhopal, India}

\begin{abstract}
In this study we derive the magnetized Taub adiabat equations from the hydrodynamic conservation conditions. We employ the magnetized taub adiabat equations to study
the evolution of a magnetized neutron star to magnetized quark star. 
The pressure of the burnt quark matter has a maximum which indicates a bound on the maximum mass of the quark star. The central density of the neutron star and the angle 
between the rotation axis and the magnetic axis (defined as the tilt angle) are seen to be significant in determining the magnetic field, and the tilt of the
quark star. The magnetic field and the tilt of the quark star can have a observational significance and can help in understanding the physics at high density and strong 
magnetic field.
\end{abstract}



\maketitle

\section{Introduction}

Shock fronts are generally depicted by a discontinuous change in the characteristics of the medium which propagates faster than the speed of sound in that medium. In most plasmas, 
the width of the shock 
front is very thin, and it is usually considered to be a one-dimensional plane of discontinuity \cite{landau}. Taub \cite{taub} was the first to study the 
relativistic hydrodynamic shocks. He used the mass, momentum, and energy conservation laws 
to derive the relativistic Rankine-Hugoniot (RH) conditions. De-hoffmann \& teller \cite{hoffmann} displayed the theoretical treatment of hydrodynamic shocks in the 
presence of a magnetic field, which was followed by an avalanche of theoretical studies \cite{landau,cabannes}. However, the relativistic treatment of magnetized 
hydrodynamic shocks was first done by Lichnerowicz \cite{lichnerowicz,webb}. Other important works in this field were successively carried out by Appl \&
Camezind \cite{appl}, Majorana \& Anile \cite{majorana} and Ballard and Heavens \cite{ballard} to name a few.

The interaction between hydrodynamic motion and magnetic fields in conducting plasmas are essential in the problem of astrophysics, 
high-energy collision, and geophysics. Two individual cases of magneto-hydrodynamic waves are common in physics; the hydrodynamic shock and the electromagnetic wave.
As the electromagnetic waves travel at the speed of light, we need to treat the problem relativistically. De-hoffmann \& Teller \cite{hoffmann} did that and treated 
the conducting fluid to be of having infinite conductivity. It was done by transforming the shock to a frame where the flow velocity is parallel to the magnetic field. 
This assumption prevents the self-induction of the magnetic field if the fluid is at rest, and is well suited for astrophysical scenarios
as the spatial dimension of most of the astrophysical problems are very large.

The standard technique of writing the jump conditions is to set the divergence of the stress-energy tensor to be zero and use the Gauss's theorem to get the jump conditions 
across the shock front. The conditions give the general mass, momentum, and energy continuity equations across the front. The three non-magnetized hydrodynamic equations 
can be expressed as a single equation known as Taub adiabat (TA) equation \cite{taub,thorne}. The equation connects the 
thermodynamic variables of one side of the front with variables on the other side, and is deprived of any velocity term. 

The astrophysical problem which frequently deals with hydrodynamic equations is the PT from a neutron star (NS) to a Quark star (QS). 
It was conjectured that at the center of NSs where supra-nuclear densities are believed to be present, normal hadronic matter (HM) is not the stable state of matter 
\cite{itoh,bodmer,witten}. At such densities 
HM is vulnerable to QM (quark being their constituent particles). Therefore, it is likely that a PT occurs taking the 
strongly interacting confined phase (HM) to a deconfined phase (QM). 
If the QM prevails over a substantial region at the core of a NS it is referred as a hybrid star and if the entire star consists of deconfined quarks it is known as a strange star.

Such PT or conversion in NS is likely to liberate a significant amount of gravitational energy, which can power gamma-ray bursts \cite{berezhiani,mallick-sahu} and 
can have gravitational wave signals \cite{mallick-mag1,mallick-apj}.
The initiation of this PT can be due to a variety of reasons: starting from cosmological quark nugget \cite{alcock}, mass-accretion \cite{alcock} to 
pulsar glitches \cite{chubarian}. The process of phase transition is also widely debated in literature: whether it is a detonation or a deflagration. One of the most usual ways
of studying the PT is by employing the hydrodynamic equations. Cho et al. \cite{cho} were probably the first ones to use the hydrodynamic jump conditions to argue that weak detonation is a possible mode of combustion. However, Tokareva et al. \cite{tokareva} and Lugones et al. \cite{lugones} argued that the possible mode of combustion could even be fast detonation.
Bhattacharyya et al. \cite{bhat,mallick-igor} argued that it can be a detonation or a deflagration depending on the density, whereas Drago et al. \cite{drago}
demonstrated that it is always a deflagration if the process is exothermic. A detailed discussion on such scenario can be found in a recent paper by 
Furusawa et al \cite{furusawa}.

Most of the studies discussed above employ the 
relativistic and non-relativistic RH equations; however, TA equation is not much studied. The magnetized counterpart of the TA is still to be derived and analyzed. In this 
article, we calculate the magnetized version of the TA equation and employ them to study the astrophysical scenario of PT.
The usefulness of this equation lies in the fact they do not involve matter velocities and only deals with the thermodynamic variables of the state like pressure, 
density and energy. 

The paper is arranged as followed: In section 2 we calculate the magnetized TA equations from the conservation equations. Next, in section 3 we discuss the EoS 
and the magnetic field configuration of the star. Section 4 is dedicated to our results, and finally, in section 5 we summarize our work and draw conclusion from them.

\section{magnetized Taub adiabat} 



The continuity equations across the front for a magnetized shock is calculated in detail by Mallick \& Schramm \cite{mallick-tl} and Mallick \& Singh \cite{mallick-mag},
and we mention only the details here.
The assumption that the fluid is infinitely conducting makes the electric field to disappear. Also, the equations are solved in a particular frame called HT frame 
\cite{hoffmann} where the magnetic field and the matter velocities are aligned. We assume that $x$-direction is normal to the shock plane. The magnetic field 
is constant and lies in the $x-y$ plane. Therefore the velocities and the magnetic fields are given by $v_x$ and $v_y$ and by 
$B_x$ and $B_y$ respectively. The angle between the magnetic field and the shock normal in the HT frame is denoted by $\theta$ ($\theta_1$ being the incidence angle and 
$\theta_2$ the reflected angle).  We assume that the PT happens as a single discontinuity separating the two phases. 
Therefore we denote $'1'$ as the initial state ahead of the shock (HM or upstream) and $'2'$ as the final state behind the shock (QM or downstream).

The Maxwell equation of no monopoles $\nabla \cdot \overrightarrow{B}=0$ gives $ B_{1x}=B_{2x}=B$, where
$B$ is the magnetic field in the x-direction, which is the same in both phases. $B_1$ and $B_2$ are magnetic fields in the y-direction in phases $1$ and $2$ respectively.
\begin{equation}\label{eq1}
\omega_{1}\gamma_{1}^{2}v_{1x}=\omega_{2}\gamma_{2}^{2}v_{2x}
\end{equation}
\begin{equation}\label{eq2}
\omega_{1}\gamma_{1}^{2}v_{1x}^{2}+p_{1}+\frac{B_{1}^{2}}{8\pi}=\omega_{2}\gamma_{2}^{2}v_{2x}^{2}+p_{2}+\frac{B_{2}^{2}}{8\pi}
\end{equation}
\begin{equation}\label{eq3}
\omega_{1}\gamma_{1}^{2}v_{1x}v_{2x}-\frac{B B_{1}}{4\pi}=\omega_{2}\gamma_{2}^{2}v_{2x} B_{2y}-\frac{B B_{2}}{4\pi}
\end{equation}
\begin{equation}\label{eq4}
n_{1}\gamma_{1}v_{1x}=n_{2}\gamma_{2}v_{2x}
\end{equation}
where, $w$ is the enthalpy ($w=\epsilon+p$), $u^{\mu}=(\gamma,\gamma v)$ is the 
normalized 4-velocity of the fluid and $\gamma$ is the Lorentz factor.
For the HT frame we also have 
\begin{eqnarray}
 \frac{v_{1y}}{v_{1x}}=\frac{B_{1}}{B}\equiv \tan\theta_1 \\
 \frac{v_{2y}}{v_{2x}}=\frac{B_{2}}{B}\equiv \tan\theta_2.
\end{eqnarray}

Therefore, we can write the y-component of velocity in terms of x-component as
\begin{equation}\label{eq5}
v_{1y}=\frac{B_1}{B}v_{1x}.
\end{equation}
The four velocity is defined as
$$ u_{1i} = v_{1i} \gamma_{1}\quad\quad\Rightarrow\quad\quad u_{1x} = v_{1x} \gamma_{1}\quad \text{and} \quad u_{1y} = v_{1y} \gamma_{1}.$$
The mass-flux $j$ in the x-direction is defined as 
\begin{equation}\label{eq6}
n_{1} u_{1x}=n_{2} u_{2x} =j
\end{equation}
or as
\begin{equation}
u_{1x}= \frac{j}{n_{1}}=jV_{1} \quad \quad \text{and} \quad \quad u_{2x}= \frac{j}{n_{2}}=jV_{2}.
\label{eq7}
\end{equation}
Using the above definitions, eqn \ref{eq1} can be redefined as 
\begin{gather*}
\omega_{1} u_{1x} \gamma_{1}=\omega_{2} u_{2x} \gamma_{2} 
\quad  \Rightarrow\quad\omega_{1} j V_{1} \gamma_{1}=\omega_{2} j V_{2} \gamma_{2} \\
\quad  \Rightarrow\quad\omega_{1}  V_{1} \gamma_{1}=\omega_{2} V_{2} \gamma_{2} 
\end{gather*}
and further defining the chemical potential as $\mu_{i} =\omega_{i} V_{i}=\omega_{i}/n_{i}$, eqn \ref{eq1} reduces to 
\begin{equation}\label{eq8}
\Rightarrow\quad\mu_{1} \gamma_{1}=\mu_{2} \gamma_{2}.
\end{equation}
  
With the definition of modified pressure $p_{i}^{'}=p_{i}+\frac{B_{i}^{2}}{8\pi}$, eqn \ref{eq2} reduces to 
\begin{eqnarray}
\omega_{1} j^{2} V_{1}^{2}+p_{1}^{'}=\omega_{2} j^{2} V_{2}^{2}+p_{2}^{'}  
\end{eqnarray}
and the square of the mass-flux $j^2$ becomes 
\begin{equation}\label{eq9}
\Rightarrow\quad  j^{2}=-\frac{(p_{2}^{'}-p_{1}^{'})}{(\mu_{2} V_{2}-\mu_{1} V_{1})}.
\end{equation}
 
Multiplying eqn \ref{eq9} by $(\mu_{1} V_{1}+\mu_{2} V_{2})$, and with the definition of $j^{2}=\frac{u_{2x}^{2}}{v_{2}^{2}}=\frac{u_{1x}^{2}}{v_{1}^{2}} $
and performing a little algebra, we have the relation
\begin{equation}
\mu_{2}^{2} u_{2x}^{2}-\mu_{1}^{2} u_{1x}^{2}=(p_{1}^{'}-p_{1}^{'})(\mu_{1} V_{1}+\mu_{2} V_{2}).  
\label{eq9-b}
\end{equation}
Squaring eqn \ref{eq8} and then subtracting eqn \ref{eq9-b} from it, we have 
\begin{equation}\label{eq10}
\mu_{2}^{2}(u_{2x}^{2}-\gamma_{2}^{2})- \mu_{1}^{2}(u_{1x}^{2}-\gamma_{1}^{2})=(p_{2}^{'}-p_{1}^{'})(\mu_{1} V_{1}+\mu_{2} V_{2})
\end{equation}
With the definition of $u_{1x}$ and $\gamma$, and adding and subtracting $\quad\frac{v_{1y}^{2}}{1-v_{1x}^{2}-v_{1y}^{2}}$, we have 
$u_{1x}^{2}-\gamma_{1}^{2}= -1-u_{1y}^{2}$. 
The above equation can also be written in terms of magnetic fields  
\begin{equation}
u_{1x}^{2}-\gamma_{1}^{2} =-1-\frac{B_{1}^{2}}{B^{2}}\, j^{2} V_{1}^{2}.
\label{eq12}
\end{equation}

Using eqn \ref{eq12}, eqn \ref{eq3} can be rewritten as 
\begin{equation}
(B_{1} \mu_{1}j^{2}V_{1}-B_{2} \mu_{2}j^{2}V_{2})=\frac{B^{2}}{4\pi}(B_{1}-B_{2}). 
\end{equation}   
From the above equation $j^2$ can also be defined as
\begin{equation}
j^{2}=\frac{B^{2}(B_{1}-B_{2})}{4\pi(B_{1} \mu_{1}V_{1}-B_{2} \mu_{2}V_{2})}
\label{eq13}
\end{equation}

Using eqn \ref{eq12} and eqn \ref{eq13}, we can write
\begin{eqnarray}
u_{2x}^{2}-\gamma_{2}^{2}=-1-\frac{B_{2}^{2}V_{2}^{2}(B_{1}-B_{2})}{4\pi(B_{1} \mu_{1}V_{1}-B_{2} \mu_{2}V_{2})}.
\label{eq15}
\end{eqnarray}

Inserting this equation in eqn \ref{eq10}, we have,
\begin{widetext}
 \begin{gather}
\mu_{2}^{2}\bigg[1+\frac{B_{2}^{2}V_{2}^{2}(B_{1}-B_{2})}{4\pi(B_{1} \mu_{1}V_{1}-B_{2} \mu_{2}V_{2})}\bigg]\mu_{1}^{2}\bigg[1+
\frac{B_{1}^{2}V_{1}^{2}(B_{1}-B_{2})}{4\pi(B_{1} \mu_{1}V_{1}-B_{2} \mu_{2}V_{2})}\bigg] 
= (p_{1}^{'}-p_{1}^{'})(\mu_{1} V_{1}+\mu_{2} V_{2}) 
\end{gather}
\end{widetext}

This is the modified "Taub Equation" for the magnetic field. Let us call it magnetized Taub adiabat 1 (MTA1) equation. The equation is independent of velocity, and 
only the thermodynamic variables and magnetic field components are present.
In the limit of no magnetic field $ B_{1}=B_{2}=0$ or even for equal magnetic field $B_{1}=B_{2}\neq 0$ this reduces to
\begin{equation}
 \mu_{2}^{2}-\mu_{1}^{2}=(p_{2}-p_{1})(\mu_{1} V_{1}+\mu_{2} V_{2})
\end{equation}
general TA.

\textbf{Some other form of Taub Adiabat}  \\

Defining $V_{i}=\frac{1}{n_{i}}$ and $X_{i}=\frac{\omega_{i}}{n_{i}^{2}}$, we have 
\begin{equation}
\omega_{2}X_{2}-\omega_{1}X_{1}= (p_{2}-p_{1}) (X_{2}-X_{1})
\end{equation}
with $j^{2}=\frac{p_{2}-p_{1}}{X_{2}-X_{1}}$. The MTA1 can also be rewritten in the given form, 
\begin{widetext}
 \begin{equation}
\boxed{\omega_{2}X_{2}\bigg[1+Y\frac{B_{2}^{2}}{n_{2}^{2}}\bigg]-\,\omega_{1}X_{1}\bigg[1+Y\frac{B_{1}^{2}}{n_{1}^{2}}\bigg] 
=(p_{2}^{'}-p_{1}^{'})(X_{2}+X_{1})}
\end{equation}
\end{widetext}
where $Y$ is defined as $Y=\frac{B_{1}-B_{2}}{4\pi(B_{1}X_{1}-B_{2}X_{2})}$.

The TA equation has two unknowns (one thermodynamic variable and one magnetic field) and only one equation, and therefore we need another equation to solve for two unknowns. 
The equation can be obtained by equating the definition of $j^2$ from eqn \ref{eq9} and eqn \ref{eq13}, 
 \begin{gather*}
-\frac{(p_{2}^{'}-p_{1}^{'})}{\bigg(\frac{\omega_{2}}{n_{2}}\cdot \frac{1}{n_{2}}-\frac{\omega_{1}}{n_{1}}\cdot\frac{1}{n_{1}}\bigg)}=\frac{B^{2}(B_{1}-B_{2})}{4\pi\bigg(B_{1} 
\frac{\omega_{1}}{n_{1}}\frac{1}{n_{1}}-B_{2} \frac{\omega_{2}}{n_{2}}\frac{1}{n_{2}}\bigg)} \\ 
\Rightarrow \quad p_{2}^{'}-p_{1}^{'}=-B^{2} Y(X_{2}-X_{1}) 
\end{gather*}
Calling the equation as MTA2, it can be further simplified as  
\begin{equation}
\boxed{( p_{2}^{'}-p_{1}^{'})(X_{1}+X_{2})=B^{2} Y(X_{1}^{2}-X_{2}^{2})}.
\end{equation}

Solving the two MTA equations (equations which are boxed), we can have all the thermodynamic properties of the downstream variables and downstream magnetic field. 
The upstream and downstream variables can be used to calculate the matter velocities of both the region. With the notation of $n,m$ and $s$ as $n=\Big(1+\frac{B_{1}}{B}\Big)$,  
$m=\Big(1+\frac{B_{2}}{B}\Big)$ and 
\begin{widetext}
\begin{equation*}
s=\sqrt{[\omega_{1}^{2}+\omega_{2}^{2}-2\omega_{1}\{\omega_{2}+2m(p_{1}-p_{2})\}+ 4\omega_{2}n(p_{1}-p_{1})+4mn(p_{1}-p_{1})^{2}]}  
\end{equation*}
the velocities can be solved in terms of thermodynamic properties. 
\begin{eqnarray}
 {v_1}^2  = \frac{\omega_{1}^{2}+2n\{\omega_{2}+m(p_{1}-p_{2})\}(p_{1}-p_{2})-\omega_{1}\{\omega_{2}+2mp_{1}+s-2mp_{2}\}}{2[\omega_{1}m-n\{\omega_{2}+m(p_{1}-p_{2})\}]
 [\omega_{1}+n(p_{2}-p_{1})]} \\ \nonumber
{v_2}^2 =\frac{\omega_{2}^{2}+\omega_{2} s-2m(p_{1}-p_{2})\{\omega_{1}+n(-p_{1}+p_{2})\}-\omega_{2}\{\omega_{1}+2n(-p_{1}+p_{2})\}}{2[-\omega_{1}m+n\{\omega_{2}+m(p_{1}-p_{2})\}]
 [\omega_{2}+m(p_{1}-p_{2})]} 
 \label{vel-1}
\end{eqnarray}
or,
\begin{eqnarray}
 {v_1}^2  = \frac{\omega_{1}^{2}+2n\{\omega_{2}+m(p_{1}-p_{2})\}(p_{1}-p_{2})+\omega_{1}\{-\omega_{2}-2mp_{1}+s+2mp_{2}\}}{2[\omega_{1}m-n\{\omega_{2}+m(p_{1}-p_{2})\}]
 [\omega_{1}+n(p_{2}-p_{1})]} \\ \nonumber
 {v_2}^2=-\frac{-\omega_{2}^{2}+\omega_{2} s+2m(p_{1}-p_{2})\{\omega_{1}+n(-p_{1}+p_{2})\}+\omega_{2}\{\omega_{1}+2n(-p_{1}+p_{2})\}}{2[-\omega_{1}m+n\{\omega_{2}
 +m(p_{1}-p_{2})\}][\omega_{2}+m(p_{1}-p_{2})]} 
 \label{vel-2}
\end{eqnarray}
\end{widetext}
\section{EoS and magnetic field}

For the hadronic phase, we adopt a relativistic mean-field approach which is generally used to describe
the HM in NSs. In our present calculation we mostly use PLZ parameter setting \cite{reinhard}, however, for some comparisons we also adopt NL3 parameter setting \cite{glendenning}.

The QM is described using MIT bag model \cite{chodos} along with the quark interaction term. The grand potential of the model is given by
\begin{equation}
\Omega_Q=\sum_i \Omega_i +\frac{\mu^4}{108\pi^2}(1-a_4)+B_g
\end{equation}
where $i$ stands for quarks and leptons, $\Omega_i$ signifies the potential for species $i$ and
$B_g$ is the bag constant with $a_4$ being the quark interaction parameter, varied between 1 (no interaction) and 0 (full interaction). 
The second term represents the interaction among quarks. We choose the values of $B_g^{1/4}=140$ MeV and $a_4=0.55$. 

In our calculation we have used a phenomenological magnetic field profile recently formulated by Dexheimer et al. \cite{veronica} where it is described as a 
function of baryon chemical potential (satisfying Einstein maxwell field equations). The magnetic field profile is given by 
\begin{equation}
 B^{*}(\mu_{B}) = \frac{a + b\mu +c{\mu_{B}}^2}{{B_c}^2} \mu_{M}.
\label{mag-prof}
\end{equation}
with coefficients $a=-0.3, b=7.0\times10^{-3}$ and $c=-1.0\times10^{-7}$. $\mu$ is given in MeV and $\mu_{M}$ the dipole magnetic moment 
in $Am^2$. The magnetic field is obtained 
in units of the critical field for the electron, $B_c = 4 . 414 \times 10^{13}$ G.

\section{Results}

The MTA equations solves for the downstream quantities treating the upstream quantities as an input. 
For non-magnetized plasma only the TA is enough, however, for magnetized plasmas, both MTA equations are required. 
First, we show a few results for the non-magnetized plasma and prove that such effects are quite general.

\begin{figure}
\includegraphics[width = 3.5in]{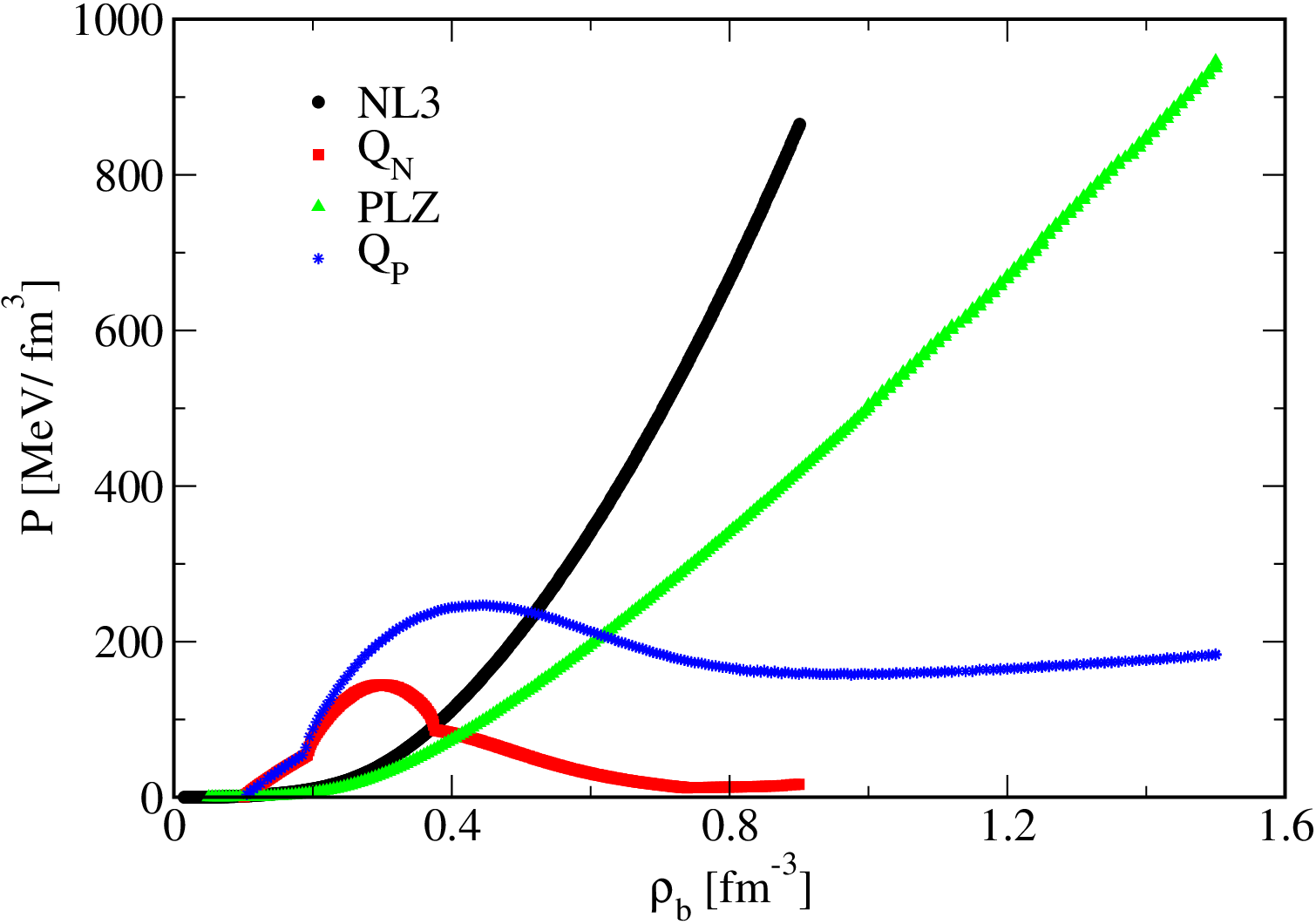} 
\caption{(Color online) $p$ as a function of $\rho_b$ for HM (NL3 (black curve) and the PLZ (green curve)) and their corresponding downstream QM (Q$_N$ and Q$_P$ marked by the red 
curve (square) and blue curve (star)) are illustrated. The burnt pressure follows a Gaussian like curve and cut their corresponding HM pressure curves at higher $\rho_b$ 
values.}
\label{fig-prs}
\end{figure}

\begin{figure}
\includegraphics[width = 3.5in]{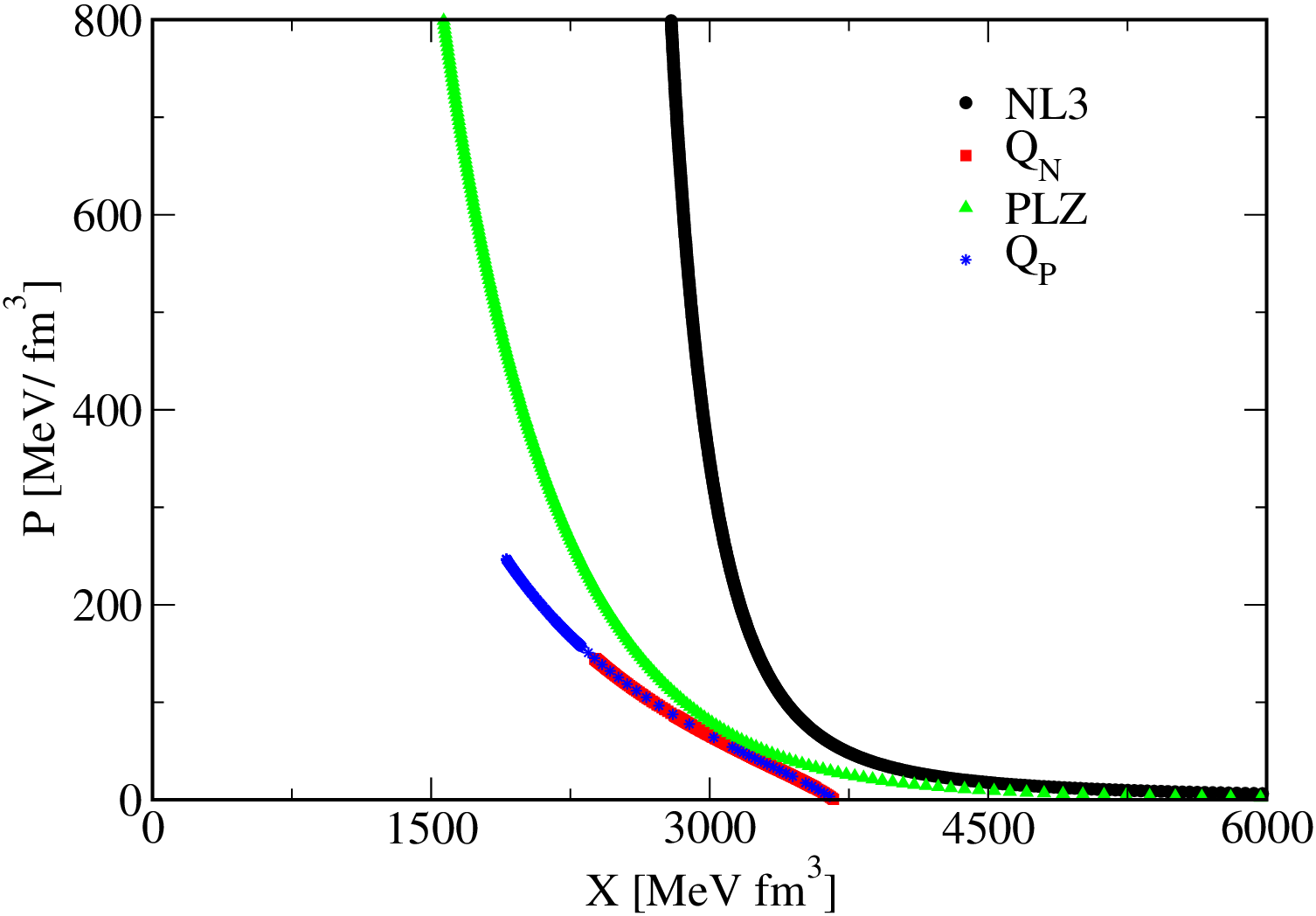} 
\caption{(Color online) The TA ($p$ vs. $X$) curves for HM (NL3 (black curve) and the PLZ (green curve)) with their corresponding burn state with Q$_N$ and Q$_P$ 
marked by the red curve (square) and blue curve (star) are drawn. The $Q_P$ curve  extends much beyond the corresponding $Q_N$ curve.
The upstream point lies on the black/green curve whereas the downstream points lie on the red/blue.}
\label{fig-taub}
\end{figure}

In fig \ref{fig-prs} we show the pressure of the upstream region and the downstream region as a function of baryon density. The HM pressure rises monotonically. 
However, the downstream or QM pressure initially increases and then comes down. It crosses the HM curve at a density higher than two-time nuclear density. 
We have shown our results for two HM EoSs. We plot the corresponding TA in fig \ref{fig-taub}. As we proceed northwards along the upstream curve treating their points 
as an input to the TA equation, the downstream curve also moves northward. However, after a certain point, although the upstream curve rises
the downstream trajectory does not rise but retraces its path. There is a maximum point for the downstream pressure. This maximum point of the downstream path coincides 
with the maximum of the pressure in fig \ref{fig-prs}.

For the MTA equations, the x-component of the 
magnetic field does not change in the HT frame and the only change is in the y-component.
The variation of the magnetic field along the star is poloidal in nature as described by eqn \ref{mag-prof}. With the given configuration, the magnetic field at the 
highest density in our calculation ($\rho_b=1.5$ fm$^{-3}$) is $1.5 \times 10^{18}$ G while at the lowest density is $4.5\times 10^{17}$ G. This variation of $B$ is constant, 
however, $B_{1}$ can change with a change in $\theta_1$. When $B > B_{1}$ the angle is small, while when $B \simeq B_{1}$ then the 
angle is close to $45^{\circ}$ and when $B < B_{1}$ the angle exceeds $45^{\circ}$.

\begin{figure}
\includegraphics[width = 3.5in]{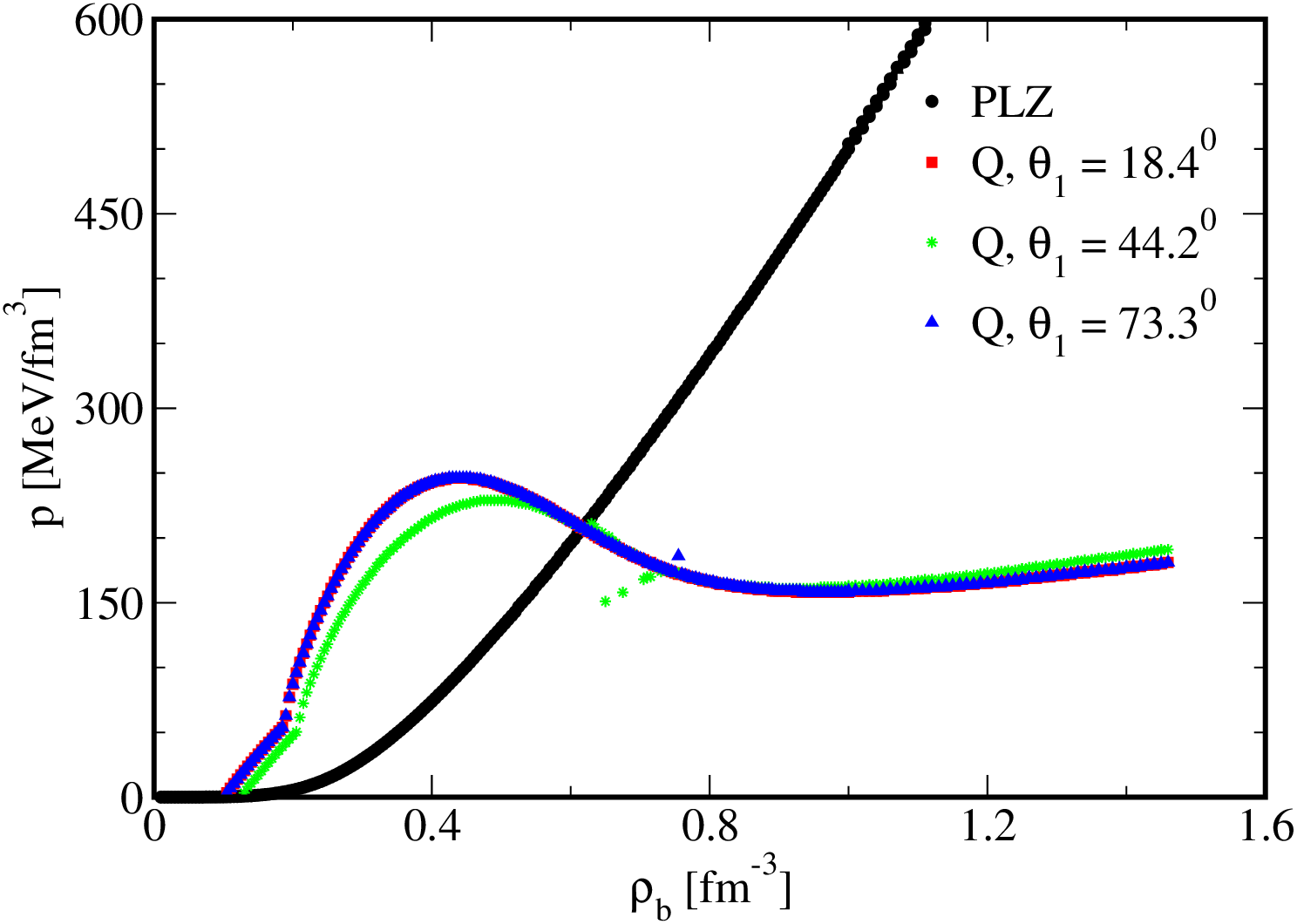} 
\caption{(Color online) $p$ as a function of $\rho_b$ for PLZ parameter and the burnt pressure of the QM is drawn. The burnt curves are drawn for 
three $\theta_1$ values. The $B$ value for the whole treatment is fixed and is $1.5\times10^{18}$ G, whereas the value of $B_{1}$ changes. 
The crossing point of the hadronic pressure and quark pressure 
is the same for all three values.}
\label{fig-prsm}
\end{figure}

\begin{figure}
\includegraphics[width = 3.5in]{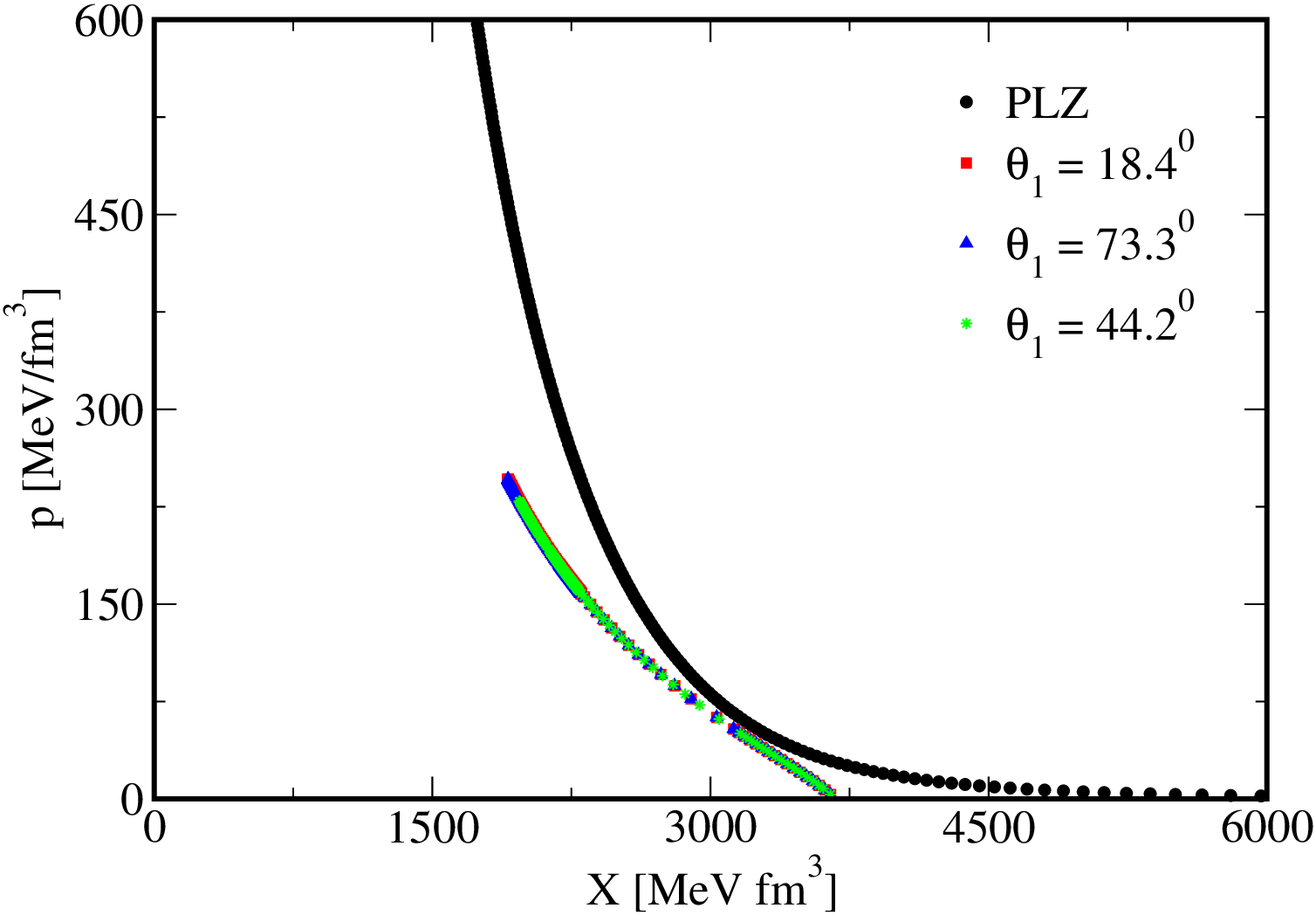} 
\caption{(Color online) The TA ($p$ vs. $X$) curves for PLZ parameter and the burnt pressure of the QM is drawn. The curve are drawn for three angular values between the 
shock front and matter velocity. The burnt curve for $\theta_1=44.2^{\circ}$ retraces from lower point than any other curve. Here curve for $\theta_1=73.3^{\circ}$ is plotted
before $\theta_1=44.2^{\circ}$ curve to emphasize that the latter curve maximum pressure value is lower than former curve.}
\label{fig-taubm}
\end{figure}

In fig \ref{fig-prsm} we plot the pressure as a function of density for three different $\theta_1$ values. For the two extreme cases, ($\theta_1 \neq 44.2^{\circ}$) the nature 
of the pressure curves are similar. However,for $\theta_1=44.2^{\circ}$, the downstream pressure is always less than the other two, though, 
the crossing of the hadronic and quark pressures coincides. Beyond that all the curve almost overlaps. We plot the TA curves in fig \ref{fig-taubm}. 
The retracing of the quark trajectory can still be seen in all the curves, however, for the curve with $\theta_1=44.2^{\circ}$ it occurs from a much lesser pressure 
value.

\begin{figure}
\includegraphics[width = 3.5in]{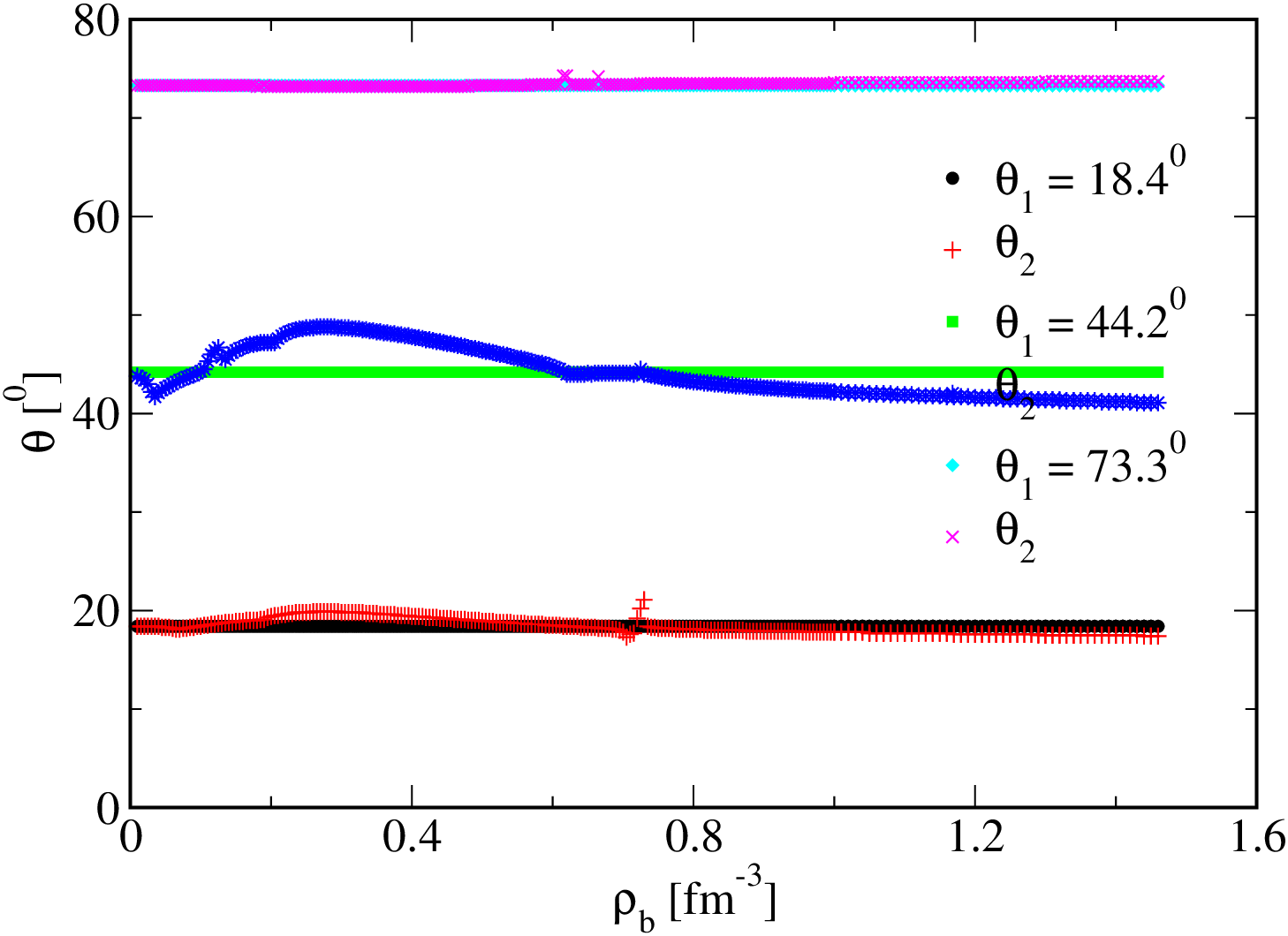} 
\caption{(Color online) The angle between downstream velocity and the shock perpendicular is shown for three different input angle $\theta_1$. The input angle 
for a particular analysis is whereas the output angle $\theta_2$ varies with density. Break in the curve reflects numerical fluctuation.}
\label{fig-theta}
\end{figure}

A change in the upstream angle imparts a change in the downstream angle as illustrated in
fig \ref{fig-theta}. For small $\theta_1$, the downstream $\theta_2$ first rises with an increase in density and reaches a maximum angle of $21^{\circ}$ and then 
decreases gradually with density finally coincideing with $\theta_1$. At further large densities $\theta_2$ becomes slightly smaller than $\theta_1$. 
For $\theta_1=44.2^{\circ}$ the initial rise in 
$\theta_2$ is much sharper and attains a maximum value of $49^{\circ}$ after which it gradually decreases. The nature of the curve is similar to that of the previous case. 
However, for large angles, $\theta_2$ is almost equal to $\theta_1$ and both the curves overlap.

\begin{figure}
\includegraphics[width = 3.5in]{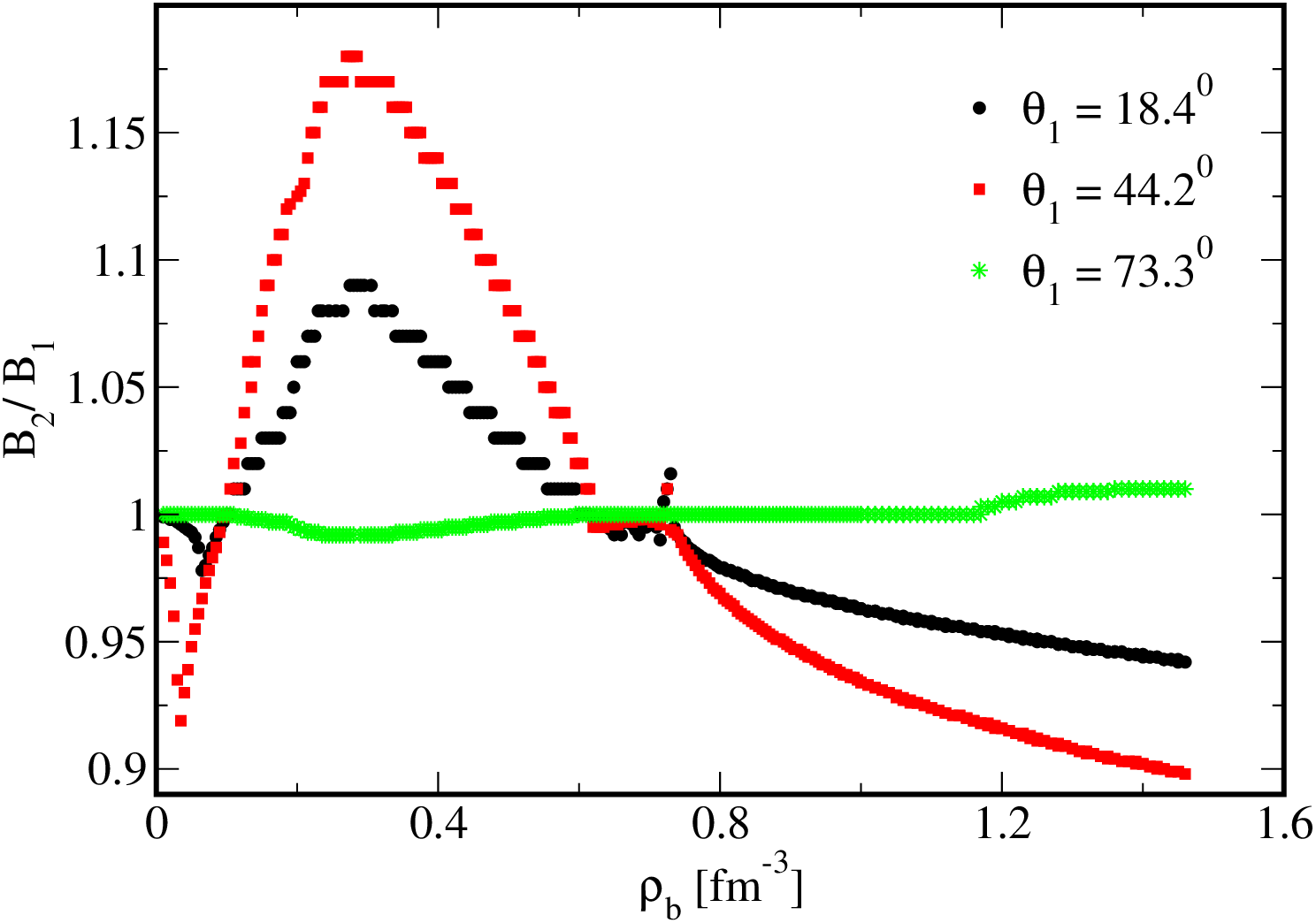} 
\caption{(Color online) The ratio of $B_{2}/B_{1}$ as a function of $\rho_b$ is shown. The ratio is shown for three input $\theta_{1}$ values, where value 1 indicating
$B_{2}=B_{1}$.}
\label{fig-mag}
\end{figure}

The nature of $\theta_1$ and $\theta_2$ is very important in determining $B_{2}$. In fig. \ref{fig-mag} we plot the ratio of $B_{2}$ to $B_{1}$ against baryon number density. 
For the case $\theta_1=18.4^{\circ}$, 
at low densities $B_{2}$ is greater than $B_{1}$ and $B_{2}$ increases with density, reaching a minimum at around $\rho_b =0.3$ fm$^{-3}$. The ratio becomes 
$1$ at $\rho_b=0.64$ fm$^{-3}$ and continues to be so for some density range.
Beyond $\rho_b=0.73$ fm$^{-3}$,  $B_{2}$ then becomes smaller than $B_{1}$. 
For $\theta_1 = 44.2^{\circ}$, at low densities $B_{2}$ is much larger than $B_{1}$ and increases further with density. The nature of the curve 
is similar to that of $\theta_1 = 18.4^{\circ}$ only differing quantitatively.  
For $\theta_1 = 73.3^{\circ}$, the nature is quite different and $B_{2}$ is almost equal to $B_{1}$ throughout the density range. 

\begin{figure}
\includegraphics[width = 3.5in]{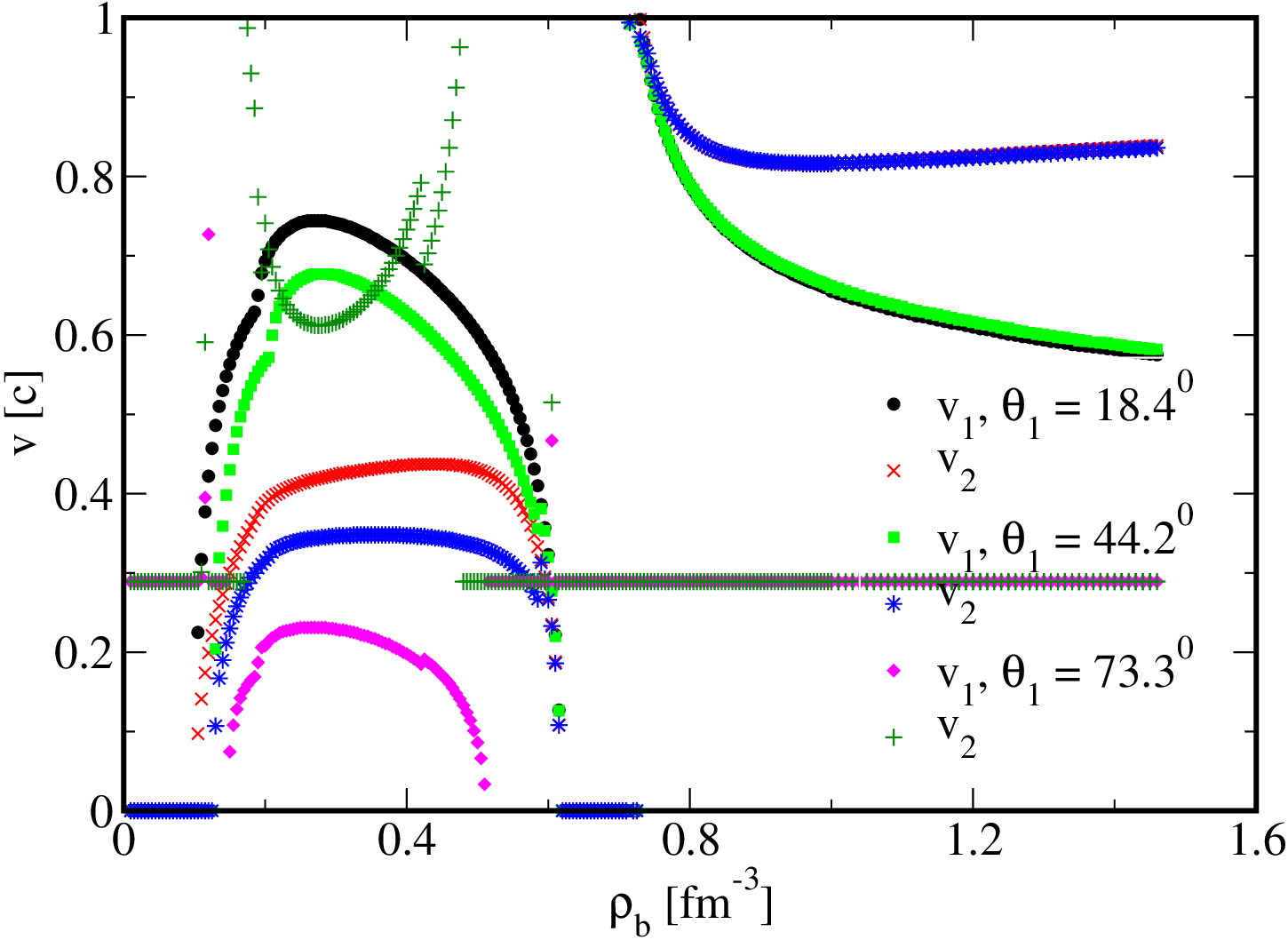} 
\caption{(Color online) The upstream ($v_1$) and downstream ($v_2$) velocities are shown as a function of $\rho_b$ for input HM EoS. Both $v_1$ and $v_2$ first increases till a 
point then decreases and goes to zero. $v_1$ is always greater than $v_2$ however they reach the maximum point and goes to zero at same $\rho_b$ values. The velocities for NL3 EoS goes to 
zero at much smaller value of $\rho_b$ than for PLZ velocities. The velocities attain high non zero values at much higher densities.}
\label{fig-vel}
\end{figure}

Solving the conservation conditions we can find the matter velocities of the upstream and downstream region in terms of the thermodynamic variables and the magnetic field as 
derived in eqn. \ref{vel-1}, \ref{vel-2}. In fig \ref{fig-vel} we have plotted the velocity as a function of density. For the first case, ($\theta_1=18.4^{\circ}$) 
both $v_1$ and $v_2$ initially increase with density, but $v_1$ is always greater than $v_2$. $v_1$ rises much faster and reaches a peak at $\rho_b=0.24$ fm$^{-3}$ beyond which
it decreases becoming zero at $\rho_b=0.74$ fm$^{-3}$. $v_2$ also increases with density (not as steeply as $v_1$) and then becomes flatter beyond 
$\rho_b=0.25$ fm$^{-3}$. The plateau region continues untill $\rho_b=0.64$ fm$^{-3}$ beyond which it falls steeply to zero. The velocities now becomes either zero or are unphysical. 
At much higher densities $v_1$ and $v_2$ again becomes finite. However, at such high densities, initially the velocity is close to $1$ and decreases with increase in density. 
In this regime $v_2$ is always greater than $v_1$.

For $\theta=44.2^{\circ}$, both $v_1$ and $v_2$ are similar to the previous curves, only differing quantitatively.
The nature of the curve for $\theta_1=73.3^{\circ}$ is quite different. At low density $v_1$ increases with density and reaches a maximum at $\rho_b \simeq 0.15$ fm$^{-3}$, 
and then gradually 
decreasing to becomes zero at $\rho_b=0.6$ fm$^{-3}$. Beyond $\rho_b=0.7$ fm$^{-3}$ $v_1$ takes a constant value of $0.29$. The $v_2$ curve is totally an inverse of the $v_1$ curve.
Initially it decreases with density from value $1$ forming an ``U'' shaped curve between $\rho_b=0.14$ fm$^{-3}$ and $\rho_b=0.53$ fm$^{-3}$. Beyond that it also assumes a constant 
value of $0.29$. In this case $v_2$ is always greater than $v_1$.

\section{Summary and conclusion}

In this article, we have derived the MTA equations, which was never realized before. These equations are different from the general conservation equations since
the velocity terms are absent. The matter velocities can be calculated from the thermodynamic variables and magnetic fields. 

QM TA shows the same retracing nature even in magnetized plasmas. The pressure curve obtained from solving the MTA equations has
a peak, which is reflected from the retracing of the TA of QM. The retracing nature of the TA and the maximum value of pressure of the QM indicates a mass bound to the QSs assumed 
to be formed by first order PT.

The angle between the matter velocities and the shock front along with the density of the NS is critical
in determining the magnetic field of the burnt star and the downstream shock velocity angle. This can have significant observational consequence as it could determine whether 
the PT to a QS would result in a star less or more magnetic than the initial NS. The initial tilt angle (angle between the rotational axis and the magnetic axis) is also the angle 
between the shock front and matter velocity, assuming that the shock spreads spherically in the star. The final tilt of the QS can be different from the initial 
inclination of the NS. For instance,
if a NS of about $1.2-1.4$ solar mass with small tilt angle suffers a PT the magnetic field of the QS would be larger than the initial NS. However, if the tilt angle is large, 
the QS has similar field strength as that of the NS. The situation is different for a more massive star, where a PT with small tilt angle would result in a QS whose magnetic field 
is less than the NS. 

The burning mechanism of the star is dependent on the magnetic field, the tilt angle and the density of the NS. A star of about $1.2-1.5$ solar mass with a moderate tilt angle 
is likely to undergo a detonation ($v_1 > v_2$) however a massive star of about $1.8-2$ solar mass with small tilt angle is expected to experience a PT via a deflagration process.
There are some stars which are not prone to PT as indicated by the zeros of the velocity curve. On the other hand, a low sized star with high tilt angle is likely to undergo a 
PT via deflagration mechanism.

To summarize, the MTA equations can be a tool to study PT in magnetized NSs. The results  can be generalized and are true for hM and QM EoS.
We are in the process of further analyzing this mechanism and studying other general features of the MTA. 

\acknowledgments
The author is grateful to the SERB, Govt. of India for monetary support in the form of Ramanujan Fellowship (SB/S2/RJN-061/2015) and Early Career Research Award (ECR/2016/000161). 
RM would also like to thank IISER Bhopal for providing all the research and infrastructure facilities.

\end{document}